\documentclass[aps,prd,twocolumn,floatfix,preprintnumbers,superscriptaddress,nofootinbib]{revtex4-2}
\usepackage{style}

\begin{document}

\title{ {\Large \bf Gravitational Waves from \textit{Type-I} Strings in a Neutrino Mass Model
}
\\}
\author{Adeela Afzal \orcidlink{0000-0003-4439-5342}}
\email{adeelaafzal@theor.jinr.ru}
\affiliation{Bogoliubov Laboratory of Theoretical Physics\\ Joint Institute for Nuclear Research\\ 141 980, Dubna, Moscow Region, Russia}

\noaffiliation

\begin{abstract}
In this work, we propose a novel realization of \textit{type-I} cosmic strings arising from the spontaneous breaking of an extended gauge symmetry $SU(2)_R\times U(1)_{B-L}$ in the context of a low-scale split seesaw mechanism for neutrino mass generation. We demonstrate that the split seesaw framework, which explains the smallness of neutrino masses, naturally motivates a small scalar self-coupling $\lambda$. This intrinsically links the neutrino mass generation mechanism to the formation of \textit{type-I} cosmic strings, where the gauge coupling dominates over the scalar self-coupling ($\beta\equiv\lambda/(2g^2)<1$). We explore the cosmological implications of these strings, including their gravitational wave signatures that are testable in current and future experiments. Our findings establish a compelling and testable connection between neutrino mass generation and cosmic string phenomenology in an underexplored region of parameter space.
\end{abstract}

\maketitle

\newpage

\section{Introduction} 
\label{sec:intro}
Understanding the origin of neutrino masses and the possible remnants of high-scale symmetry breaking in the early universe are central challenges in modern particle physics and cosmology. Left-right (LR) symmetric models, based on the gauge group
$$SU(3)_C \times SU(2)_L \times SU(2)_R \times U(1)_{B-L},$$
offer a compelling framework to address these questions. They naturally incorporate right-handed (RH) neutrinos as gauge singlets under the Standard Model, but charged under $SU(2)_R \times U(1)_{B-L}$, and thus enable the implementation of the seesaw mechanism for neutrino mass generation \cite{Mohapatra1992, PhysRevLett.44.912}. Moreover, they provide a natural origin for parity violation in weak interactions, embedding it within a spontaneously broken symmetry \cite{Senjanovic:2019moe}. The leptogenesis in the LR model has been discussed in Ref.~\cite{Khalil:2012nd} and some analysis of GW production in the LR model has been discussed in Ref.~\cite{Maji:2023fhv,Ahmad:2025dds, Ahmed:2025crx, Afzal:2023kqs}. The study of leptogenesis and GWs from $U(1)_{B-L}$ breaking has been discussed in \cite{Blasi:2020wpy}, \textit{type-1} CSs in MSSM connecting electron neutrino asymmetry hinted by the
 recent EMPRESS data have recently been explored in \cite{Bae:2025gpc}.

Among the various implementations of the seesaw mechanism, the split seesaw scenario \cite{Kusenko:2010ik,Adulpravitchai:2011rq} provides an elegant explanation for the tiny active neutrino masses through geometric suppression in higher-dimensional theories. In this framework, RH neutrinos are localized at different positions in a compactified extra dimension. The overlap of their wavefunctions with the Standard Model brane results in exponentially suppressed Yukawa couplings, without requiring ultra-small fundamental parameters. This separation not only generates a realistic neutrino mass spectrum but also enables one of the RH neutrinos to be long-lived and a viable dark matter candidate \cite{Asaka:2005pn,Kadota:2007mv}.

A critical prediction of LR symmetric theories is the spontaneous breaking of the gauge symmetry $SU(2)_R\times U(1)_{B-L}\rightarrow U(1)_Y$, which is typically achieved by the vacuum expectation value (VEV) of a RH triplet scalar $\Delta_R$. This symmetry breaking can lead to the formation of topological defects, specifically cosmic strings (CSs), if the vacuum manifold contains non-trivial first homotopy groups \cite{Vilenkin:2000jqa, Kibble:1976sj}. In our model, this breaking gives rise to \textit{type-I} CSs, characterized by a scalar self-coupling $\lambda$ much smaller than the gauge coupling squared $g^2$, i.e., $\beta\equiv\lambda/(2g^2)\ll1$ \cite{Hindmarsh:1994re}. The formation of \textit{type-I} CSs from supersymmetric flat directions has been explored in \cite{Cui:2007js}. Such strings are attractive in nature and can influence early-universe dynamics and gravitational wave (GW) backgrounds.

Furthermore, the large VEV, $v_R$ associated with $\Delta_R$ not only sets the symmetry-breaking scale but also giving mass to the heavy $U(1)_{B-L}$ gauge boson $Z^{'}$. In the split seesaw context, we show that the scale $v_R\simeq(10^5-10^{15})$~GeV that is consistent with both GW observations and evades the collider bounds \cite{Queiroz:2024ipo}. The heavy mass of the gauge boson suppresses its emission from the string network, favoring decay channels such as GWs or sterile neutrino emission. This suppression is essential in our model, where we explore the cosmological signatures of such strings without conflicting with astrophysical constraints. 

In this work, we analyze the formation and evolution of \textit{type-I} CSs in the framework of the split seesaw mechanism embedded in an LR symmetric model. We discuss the implications for neutrino mass generation and CS phenomenology and identify benchmark parameter sets consistent with current theoretical and observational bounds. Our results demonstrate the compatibility of this unified framework with high-energy particle physics and early universe cosmology. This regime is often overlooked in the literature, as most cosmic string studies focus on \textit{type-II} strings, where parallel strings tend to repel at large distances while antiparallel strings attract \cite{Vilenkin:2000jqa}.

The unique synergy of this framework is threefold: Firstly, the split seesaw provides a natural explanation for small neutrino masses and motivates the small quartic coupling necessary for \textit{type-I} strings. Secondly, the left-right symmetry-breaking scale $v_R$ is simultaneously the source of heavy neutrino masses and the string tension. Thirdly, this connects the scale of neutrino mass generation directly to an observable cosmological signal-GWs from CSs—since we have shown that the mass suppression factor is constrained by the choice of $\beta$.

The paper is organized as follows: in Sec.~\ref{sec:CSLRmodel}, we discuss the LR symmetric model and the formation of CSs as a result of symmetry breaking; we define the Lagrangian and scalar potential in Sec.~\ref{sec:LagrScalarPoten}, in Sec.~\ref{sec:massgeneration} we discuss the neutrino mass generation in the split seesaw framework. We briefly discuss \textit{type-1} CSs and their phenomenological implications in Sec.~\ref{sec:type1cs} and lastly, in Sec.~\ref{Con} we summarize our conclusions.
\section{Cosmic Strings in a Left-Right Symmetric Model}
\label{sec:CSLRmodel}
We consider a left-right (LR) symmetric extension of the Standard Model based on the gauge group:
\begin{equation}
G_{\text{LR}} = SU(3)_C \times SU(2)_L \times SU(2)_R \times U(1)_{B-L},
\end{equation}
which naturally incorporates the seesaw mechanism for neutrino mass generation. The field content of the LR model is defined in Tab.~\ref{tab:lrsm-fields}
\begin{table}[tbh]
\centering
\resizebox{0.9\columnwidth}{!}{%
\begin{tabular}{|c|c|l|}
\hline
\textbf{Field} & \textbf{Symbol} & \textbf{Representation} \\
\hline
LH quark doublet & $Q_L$ & $(\mathbf{3}, \mathbf{2}, \mathbf{1}, +\tfrac{1}{3})$ \\
RH quark doublet & $Q_R$ & $(\mathbf{3}, \mathbf{1}, \mathbf{2}, +\tfrac{1}{3})$ \\
LH lepton doublet & $L_L$ & $(\mathbf{1}, \mathbf{2}, \mathbf{1}, -1)$ \\
RH lepton doublet & $L_R$ & $(\mathbf{1}, \mathbf{1}, \mathbf{2}, -1)$ \\
\hline
Higgs bidoublet & $\Phi$ & $(\mathbf{1}, \mathbf{2}, \mathbf{2}, 0)$ \\
Left Higgs triplet & $\Delta_L$ & $(\mathbf{1}, \mathbf{3}, \mathbf{1}, +2)$ \\
Right Higgs triplet & $\Delta_R$ & $(\mathbf{1}, \mathbf{1}, \mathbf{3}, +2)$ \\
\hline
\end{tabular}
}
\caption{Field content of the minimal left-right symmetric model with representations under $SU(3)_C \times SU(2)_L \times SU(2)_R \times U(1)_{B-L}$.}
\label{tab:lrsm-fields}
\end{table}
The symmetry is broken by a RH scalar triplet:
\begin{equation}
\Delta_R \sim (1, 1, 3, +2),
\end{equation}
which acquires a VEV, spontaneously breaking the gauge group via:
\begin{equation}
SU(2)_R \times U(1)_{B-L} \xrightarrow{\langle \Delta_R \rangle} U(1)_Y.
\end{equation}
This breaking leads to a non-trivial vacuum manifold
\begin{equation}
\mathcal{M} = \frac{SU(2)_R \times U(1)_{B-L}}{U(1)_Y},
\end{equation}
with the first homotopy group
\begin{equation}
\pi_1(\mathcal{M}) = \mathbb{Z},
\end{equation}
indicating the formation of topologically stable CSs. For a detailed discussion on the formation of CSs, we refer the reader to \cite{Kibble:1982ae, Jeannerot:2003qv}.

\subsection{Lagrangian and Scalar Potential}
\label{sec:LagrScalarPoten}
The relevant kinetic and potential terms for $\Delta_R$ are given by:
\begin{equation}
\label{eqn:RLag}
\mathcal{L} \supset \text{Tr}[(D_A \Delta_R)^\dagger (D^A \Delta_R)] - V(\Delta_R),
\end{equation}
with the covariant derivative defined as:
\begin{equation}
\label{eqn:Rcovderi}
D_A \Delta_R = \partial_A \Delta_R - i g\, [W_{RA}, \Delta_R] - i 2 g_{B-L} B_A \Delta_R,
\end{equation}
where $\partial_A$ is the ordinary derivative, $g,\, W_{RA}$ is the associated gauge coupling and bosons with $SU(2)_R$ respectively, $g_{B-L},\, B_{A}$ is the associated gauge coupling and bosons of $U(1)_{B-L}$.
The scalar potential for $\Delta_R$ is,
\begin{equation}
\label{eqn:Rpoten}
V(\Delta_R) = -\mu^2 \, \text{Tr}[\Delta_R^\dagger \Delta_R] + \lambda \left( \text{Tr}[\Delta_R^\dagger \Delta_R] \right)^2.
\end{equation}
Here $\mu$ is a positive mass-squared parameter that triggers spontaneous symmetry breaking when the potential is minimized and $\lambda$ is the quartic self-coupling constant.
A complete treatment of the scalar potential in the LRSM requires including cross-couplings between $\Phi$, $\Delta_L$, and $\Delta_R$ to ensure the desired vacuum alignment. A general renormalizable scalar potential in the minimal LRSM can be written as \cite{BhupalDev:2018xya}
\begin{widetext}
\begin{eqnarray}
V & \ = \  & -\mu_{1}^{2}\text{Tr}[\Phi^{\dagger}\Phi]-\mu_{2}^{2}\left(\text{Tr}[\tilde{\Phi}\Phi^{\dagger}]+\text{Tr}[\tilde{\Phi}^{\dagger}\Phi]\right)-\mu_{3}^{2}\left(\text{Tr}[\Delta_{L}\Delta_{L}^{\dagger}]+\text{Tr}[\Delta_{R}\Delta_{R}^{\dagger}]\right)+\lambda_{1}\text{Tr}[\Phi^{\dagger}\Phi]^{2}\nonumber \\
 &  & +\lambda_{2}\left(\text{Tr}[\tilde{\Phi}\Phi^{\dagger}]^{2}+\text{Tr}[\tilde{\Phi}^{\dagger}\Phi]^{2}\right)+\lambda_{3}\text{Tr}[\tilde{\Phi}\Phi^{\dagger}]\text{Tr}[\tilde{\Phi}^{\dagger}\Phi]+\lambda_{4}\text{Tr}[\Phi^{\dagger}\Phi]\left(\text{Tr}[\tilde{\Phi}\Phi^{\dagger}]+\text{Tr}[\tilde{\Phi}^{\dagger}\Phi]\right)\nonumber \\
 &  & +\rho_{1}\left(\text{Tr}[\Delta_{L}\Delta_{L}^{\dagger}]^{2}+\text{Tr}[\Delta_{R}\Delta_{R}^{\dagger}]^{2}\right)+\rho_{2}\left(\text{Tr}[\Delta_{L}\Delta_{L}]\text{Tr}[\Delta_{L}^{\dagger}\Delta_{L}^{\dagger}]+\text{Tr}[\Delta_{R}\Delta_{R}]\text{Tr}[\Delta_{R}^{\dagger}\Delta_{R}^{\dagger}]\right)\nonumber \\
 &  & +\rho_{3}\text{Tr}[\Delta_{L}\Delta_{L}^{\dagger}]\text{Tr}[\Delta_{R}\Delta_{R}^{\dagger}]+\rho_{4}\left(\text{Tr}[\Delta_{L}\Delta_{L}]\text{Tr}[\Delta_{R}^{\dagger}\Delta_{R}^{\dagger}]+\text{Tr}[\Delta_{L}^{\dagger}\Delta_{L}^{\dagger}]\text{Tr}[\Delta_{R}\Delta_{R}]\right) \\
 &  & +\alpha_{1}\text{Tr}[\Phi^{\dagger}\Phi]\left(\text{Tr}[\Delta_{L}\Delta_{L}^{\dagger}]+\text{Tr}[\Delta_{R}\Delta_{R}^{\dagger}]\right)+\alpha_{3}\left(\text{Tr}[\Phi\Phi^{\dagger}\Delta_{L}\Delta_{L}^{\dagger}]+\text{Tr}[\Phi^{\dagger}\Phi\Delta_{R}\Delta_{R}^{\dagger}]\right)\nonumber \\
 &  & +\alpha_{2}\left(\text{Tr}[\Delta_{L}\Delta_{L}^{\dagger}]\text{Tr}[\tilde{\Phi}\Phi^{\dagger}]+\text{Tr}[\Delta_{R}\Delta_{R}^{\dagger}]\text{Tr}[\tilde{\Phi}^{\dagger}\Phi]+{\rm h.c.}\right) \nonumber \\
 &  & +\beta_{1}\left(\text{Tr}[\Phi\Delta_{R}\Phi^{\dagger}\Delta_{L}^{\dagger}]+\text{Tr}[\Phi^{\dagger}\Delta_{L}\Phi\Delta_{R}^{\dagger}]\right)+\beta_{2}\left(\text{Tr}[\tilde{\Phi}\Delta_{R}\Phi^{\dagger}\Delta_{L}^{\dagger}]+\text{Tr}[\tilde{\Phi}^{\dagger}\Delta_{L}\Phi\Delta_{R}^{\dagger}]\right)\nonumber \\
 &  & +\beta_{3}\left(\text{Tr}[\Phi\Delta_{R}\text{\ensuremath{\tilde{\Phi}^{\dagger}\Delta_{L}^{\dagger}}}]+\text{Tr}[\Phi^{\dagger}\Delta_{L}\text{\ensuremath{\tilde{\Phi}\Delta_{R}^{\dagger}}}]\right) \nonumber \,,\label{eq:LRV-4}
\end{eqnarray}
\end{widetext}
where $\tilde{\Phi} = \sigma_2 \Phi^* \sigma_2$ (with $\sigma_2$ the second Pauli matrix). The vacuum alignment relevant for our analysis is
\begin{align}
\langle\Phi\rangle=\begin{pmatrix}
\kappa_1 & 0 \\
0 & \kappa_2
\end{pmatrix},\quad \langle\Delta_L\rangle=0,\quad\langle\Delta_R\rangle=\dfrac{1}{\sqrt{2}}
\begin{pmatrix}
0 & 0 \\
v_R & 0
\end{pmatrix},
\end{align}
with $v_R\gg\kappa_1$, $\kappa_2\sim v_\text{EW}$.  This structure arises in the type‑I seesaw limit of the LRSM, where the cross‑couplings between $\Phi$, $\Delta_L$ and $\Delta_R$ in the full scalar potential can be chosen small enough that the global minimum satisfies $\langle \Delta_L \rangle \approx 0$, $\langle\Phi\rangle\simeq v_\text{EW} \ll \langle \Delta_R \rangle \simeq v_R$ while maintaining vacuum stability~\cite{Mohapatra:1979ia}.  Any induced $\langle\Delta_L\rangle$ is then suppressed as $v_L\sim v_\text{EW}^2/v_R$ and is negligible for neutrino masses. Crucially, the parameter $\beta=\lambda/(2g^2)$ depends only on the self‑coupling $\lambda\equiv\rho_1$ of $\Delta_R$ and the gauge coupling $g$. The LRSM features a two-stage symmetry breaking:
 \begin{align}
 \label{eqn:LRSMbreaking}
 SU(2)_L \times SU(2)_R \times U(1)_{B-L} &\xrightarrow[]{\langle \Delta_R \rangle} SU(2)_L \times U(1)_Y\\\notag
 &\xrightarrow[]{\langle \Phi \rangle} U(1)_\text{EM}.
 \end{align}
 For the purpose of studying CS formation from the primary symmetry breaking $SU(2)_R \times U(1)_{B-L} \to U(1)_Y$, we work in the limit where $v_R \gg v_{\text{EW}}$. This hierarchy is phenomenologically required and theoretically well-motivated in the standard seesaw mechanism. However, in our framework, neutrino masses are generated via the split seesaw mechanism and to be in the observable regime
 as shown in Fig.~\ref{fig:massessymmbreaking}, our analysis considers $v_R$ in the range $10^5 - 10^{15}$ GeV, still maintains $v_R \gg v_\text{EW}$. The influence of $\Phi$ and $\Delta_L$ on these specific string characteristics is subleading, as their VEVs are much smaller than $v_R$. The first breaking in Eq.~\ref{eqn:LRSMbreaking} produces Majorana masses for the RH neutrinos, a key element of the seesaw framework \cite{Mohapatra:1974gc, Senjanovic:1975rk, Mohapatra:1979ia}.
 
Assuming a vacuum configuration:
\begin{equation}
\langle \Delta_R \rangle = \frac{1}{\sqrt{2}} \begin{pmatrix}
0 & 0 \\
v_R & 0
\end{pmatrix},
\end{equation}
 which is a triplet under the gauge group $SU(2)_R$. This triplet is used to break the RH gauge symmetry $SU(2)_R \times U(1)_{B-L}$ down to the Standard Model’s hypercharge group $U(1)_Y$.
 This breaking gives rise to the formation of CSs. In the Abelian-Higgs model, the string tension is defined to be
 \begin{align}
 \label{Stringtensnu}
     \mu_{cs}\simeq 2 \pi B(\beta)v_R^2.
 \end{align}
Here $v_R$ is the VEV of the scalar field $\Delta_R$, $\beta\equiv m_{\Delta_R}^2/m_{W_R}^2$ is the ratio between the mass square of the scalar and gauge field.
The function $B(\beta)$ encodes the dependence on the scalar-to-gauge mass ratio. While its asymptotic behavior is well-known  ($B(\beta) \sim (ln(1/\beta))^{-1}$ for $\beta \ll 1$,
$B(\beta) \rightarrow 1$ for $\beta = 1$ and $B(\beta) \sim ln(\beta)$ for $\beta \gg 1$) \cite{Yung:1999du,Shifman:2002yi}, we adopt a phenomenological interpolation formula that approximates numerical solutions across $\beta\leq 1$: 
\begin{align}
\label{betaeqn}
    B(\beta)\simeq (1 + |Log\left(\beta\right)|)^{-1}.
\end{align}
This fit is chosen for its simplicity and accuracy in representing the region of our interest ($\beta \leq 1$). This fit reproduces the asymptotic behavior for $\beta\ll1$ given in \cite{Yung:1999du,Shifman:2002yi} to within $5\%$. We have shown the impact of $\beta$ on the CSs in Fig.~\ref{fig:beta_var}.
\subsection{Mass Generation in Split Seesaw}
\label{sec:massgeneration}
In the minimal LRSM, the neutrino mass matrix arises from the Yukawa interactions and the seesaw mechanism. The relevant Yukawa Lagrangian terms are:
\begin{align}
\mathcal{L}_Y \supset -Y_\ell \overline{L}_L \Phi L_R - Y_D \overline{L}_L \tilde{\Phi} L_R - \frac{1}{2} Y_R \overline{L_R^c} \Delta_R L_R + \text{h.c.}
\end{align}
 After symmetry breaking the $6\times6$ neutrino mass matrix in the basis $(v_L,v_R)$ can be written as
\begin{align}
\label{eqn:LRSM_seesaw_Massmatrix}
M_\nu = \begin{pmatrix}
m_L & m_D \\
m_D^T & M_R
\end{pmatrix},
\end{align}
with $m_L=Y_L v_L, m_D = Y_D v_\text{EW}$ and $M_R = Y_R v_R$. Diagonalizing this matrix gives
\begin{align}
\label{eqn:diagnal_massmatrix}
    M_\nu=m_L-m_D\,M_R^{-1}\,m_D^T.
\end{align}
In the minimal LRSM and also in our case as discussed in Sec.~\ref{sec:LagrScalarPoten}, $v_L$ is negligible and therefore Eq.~\ref{eqn:diagnal_massmatrix} reduces to the standard type-I seesaw formula
\begin{align}
\label{eqn:seesawType1_mass}
    M_\nu=-m_D\,M_R^{-1}\,m_D^T.
\end{align}
However, in the full LRSM, the LH triplet $\Delta_L$ may also acquire a small induced VEV, $v_L$, giving an additional contribution $m_L=Y_L v_L$ to the neutrino mass matrix in Eq.~\ref{eqn:diagnal_massmatrix}. The split seesaw mechanism \textemdash where the smallness of the neutrino Yukawa coupling arises from localization in extra dimensions \cite{Kusenko:2010ik, Adulpravitchai:2011rq} \textemdash can be consistently formulated including this term. We assume a fifth dimension $y \in [0, L]$ with the Standard Model fields (including $L_L,\Phi,\Delta_L$) are localized on a 4D brane at $y = 0$. The RH neutrinos $N_i$ and the scalar triplet $\Delta_R$ propagate in the Bulk. Their zero mode wavefunctions are chosen to be localized near the distant brane at $y=L$ in order to suppress couplings to the SM brane. For the leptonic sector, the relevant Yukawa interactions and the $5$D kinetic term for the RH neutrinos are
\begin{align}
    S^{5D}_\text{lep}&=\int d^4x \int^L_0dy \left[\delta(y)\left(-Y_\ell\bar{L}_L\Phi L_R-Y_D\bar{L}_L\tilde{\Phi}L_R\right.\right. \\\notag
     &\left.\left.-\dfrac{1}{2}Y_L\bar{L}_L^c \Delta_L L_L+\text{h.c}\right)+\left(\bar{N} i \Gamma^A D_A N -M_i \bar{N}N+\text{h.c}\right)\right. \\\notag
     &\left.+\left(Y_R\bar{N^c}\Delta_RN+\text{h.c}\right)
     \right],
\end{align}
where $N$ is the $5$D fermion containing the RH neutrino zero mode, $M_i$ is its bulk mass assumed to be positive, $Y_R$ is the $5$D Yukawa coupling, $D_A=\partial_A-ig W^a_{RA}T^a-i\,2\,g_{B-L}B_A$  ($A=0,1,2,3,4$ and $T^a$ are the generators of $SU(2)_R$) is the $5$D covariant derivative and $\Gamma^A=(\gamma^\mu,i\gamma^5)$ are the $5$D gamma matrices. The $5$D wavefunction of the $i$-th RH neutrino, localized near the brane at $y=L$, takes the form:
\begin{equation}
\Psi_{N_i}(x, y) = \psi_{N_i}(x) f_i(y), \quad \text{with} \quad f_i(y) \simeq \sqrt{2 M_i}\, e^{-M_i |y-L|}.
\end{equation}
Similarly, the bulk scalar field $\Delta_R$ has a wavefunction profile
\begin{align}
    \Delta_R(x,y)=\Phi_{\Delta_R}(x) f_\Delta(y), \quad \quad f_\Delta(y) \simeq \sqrt{2 M_\Delta}\, e^{-M_\Delta |y-L|},
\end{align}
where $M_\Delta>0$ localizes $f_\Delta(y)$ away from $y=0$. 
The 5D Yukawa coupling to the Higgs field $\Phi$ localized at $y = 0$ generates a suppressed effective 4D Yukawa coupling:
\begin{equation}
\label{5DYukawaSupression}
Y^{\text{eff}}_{i\alpha} = Y^{(5D)}_{i\alpha} \cdot f_i(0) \propto Y^{(5D)}_{i\alpha} \cdot e^{-M_i L}.
\end{equation}
After dimensional reduction, the effective $4$D Dirac and Majorana masses are
 \begin{align}
 \label{eqn:Diracmasssuppress}
  m_D &= Y^{\text{eff}}_{D} v_\text{EW}\sim  e^{- M_iL}v_\text{EW}.
 \end{align}
 The Majorana mass arises from a coupling of $\Psi_{N_i}$ to the $\Delta_R$ field in the bulk. For the Yukawa interaction, the wavefunction overlap integral is dominated by the region near $y=L$, giving
\begin{equation}
M_{N_i} \sim v_R.
\end{equation}
 where $v_R=\langle\Phi_{\Delta_R}\rangle$ is the $4$D VEV. We assume that $\Delta_L$ is localized at $y=0$, therefore, $m_L=Y_L v_L$. The neutrino mass from Eq.~\ref{eqn:diagnal_massmatrix} is
 \begin{align}
     M_\nu\simeq Y_L v_L+e^{-2M_iL}\dfrac{v_\text{EW}^2}{v_R}.
 \end{align}
For phenomenological consistency of the LRSM (electroweak precision, neutrino masses, and collider bounds) requires the hierarchy $v_R\gg \kappa_{1,2}\gg v_L$ \cite{PhysRevD.44.837}. 
In our split‑seesaw realization, $v_R$ is the high‑scale VEV of $\Delta_L$ localized on a distant brane, while $v_L$
 is suppressed both by the LRSM scalar potential (``VEV seesaw" \cite{PhysRevD.44.837}) and by the spatial separation of $\Delta_L$ from the symmetry-breaking sector. Hence, working in the limit 
$v_L=0$ is well-motivated and does not introduce fine-tuning. Without the loss of generality, the standard type-I seesaw formula, Eq.~\ref{eqn:seesawType1_mass} then gives:
\begin{equation}
\label{eqn:seesawmassformula}
M_\nu \simeq - m_D M_N^{-1} m_D^T \sim \left( e^{-M_i L} \right)^2 \cdot \frac{v_{\text{EW}}^2}{v_R},
\end{equation}
naturally yielding small active neutrino masses without requiring tiny dimensionless couplings. 
Moreover, in a consistent extra-dimensional framework, the field $\Delta_R$ is bulk. Its localization in the extra dimension leads to a universal exponential suppression of all its $4$D effective couplings, including its quartic self-coupling. The exponential suppression of the effective quartic coupling $\lambda$ arises from the volume suppression of the $5$D interaction. Following Refs.~\cite{Kusenko:2010ik}, if the $\Delta_R$ field is a bulk field with a wavefunction $f_{\Delta_R}(y)\propto e^{-M_\Delta y}$, the effective $4$D coupling is given by:
  \begin{align}
      \lambda\equiv\lambda_{4D}=\lambda_5 \int_0^L dy\,|f_{\Delta_R}(y)|^4 \propto \lambda_5\,e^{-4M_\Delta L}
  \end{align}
  where $\lambda_5$ is the fundamental $5$D coupling. This integral is the consequence of integrating the local $5$D interaction over the extra dimension. Therefore, a large $M_\Delta L$ leads to $\lambda_{4D}\ll1$, satisfying the condition for \textit{type-I} strings.
  
The exponential suppression $e^{-M_i L}$ requires $L > 1/M_i$ to be significant. However, to ensure the validity of the $4$D effective field theory (EFT) at the temperature of the $SU(2)_R \times U(1)_{B-L}$ phase transition ($T \sim v_R$), we must impose that the compactification scale, $M_c \sim 1/L$ is significantly higher than $v_R$. This avoids Kaluza-Klein mode excitations and higher-dimensional operators disrupting the dynamics of CS formation. Consequently, we require
\begin{align}
    v_R \ll M_c < M_i.
\end{align}
This hierarchy ensures that the $4$D description of the phase transition and the resulting topological defects remains robust. The small Yukawa couplings are generated by the large bulk mass terms $M_i$, not an excessively large extra dimension that would compromise the EFT.

The split seesaw mechanism accommodates the full neutrino flavor structure through the introduction of three RH neutrinos with different bulk mass parameters $M_i (i = 1,2,3)$, as shown in Ref.~\cite{Kusenko:2010ik,Adulpravitchai:2011rq}. Each RH neutrino has a hierarchically exponentially suppressed $e^{-M_iL}$  wavefunction that generates the required mass-squared differences $\Delta m_{21}^2$ and $|\Delta m_{31}|^2$ which accommodates the observed neutrino oscillation, while the structure of the 5D Yukawa matrix $Y_{i\alpha}^{(5D)}$ determines the PMNS mixing angles. As demonstrated in Ref.~\cite{Adulpravitchai:2011rq}, this framework can successfully reproduce neutrino oscillation data when implemented with flavor structures. While our CS analysis (Sec.~\ref{sec:type1cs}) considers a common breaking scale, $v_R$, the presence of multiple RH neutrinos with different flavors $M_i$ does not affect the string formation since the string tension is flavor blind. The topological defect formation depends only on the gauge symmetry breaking pattern and vacuum manifold. Although different $M_i$ could lead to varied particle emission channels, in our parameter space the dominant energy loss mechanism is gravitational radiation, which remains flavor-independent.

\begin{figure}[t]
    \centering
    \includegraphics[width=\columnwidth]{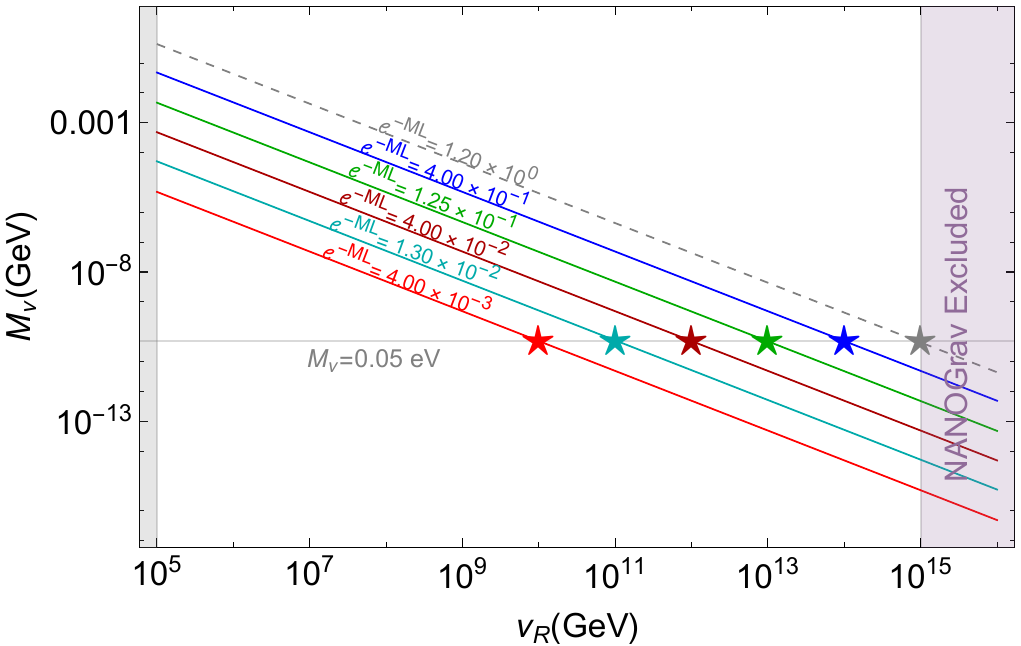}
    \caption{Neutrino mass evolution versus symmetry-breaking scale has been presented. The colored shaded region indicates the NANOGrav \cite{NANOGrav:2023gor} excluded regime as shown in Fig.~\ref{fig:omegaftype1}. The colored curves represent the different values of the exponential suppression factor. The colored stars correspond to the atmospheric scale mass, $M_\nu\simeq0.05$~eV for each exponential suppression factor. The colliders constrain the lower value $v_R\gtrsim10^5$~GeV \cite{Queiroz:2024ipo}, shown as the gray shaded regime.}
    \label{fig:massessymmbreaking}
\end{figure}

\begin{figure}[t]
    \centering
    \includegraphics[width=\columnwidth]{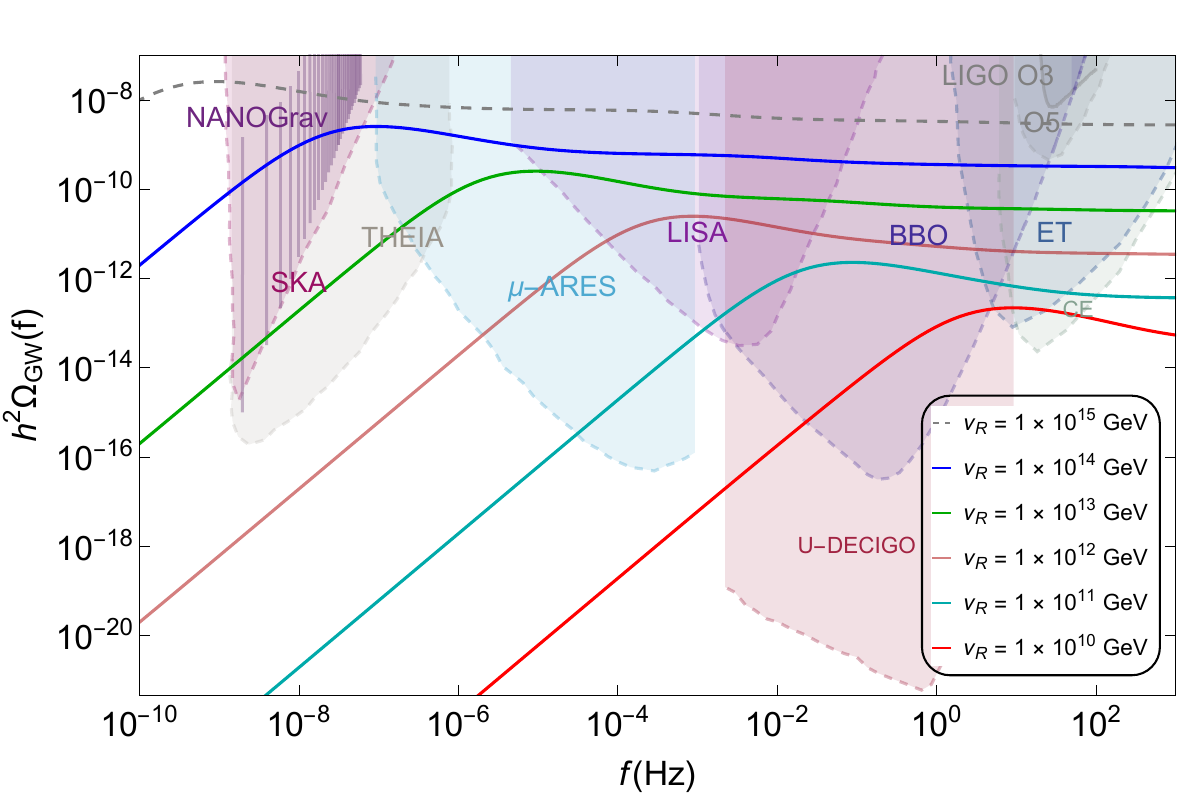}
    \caption{The evolution of GW spectra for \textit{type-I} CSs for different values of $v_R$ corresponding to the benchmark points given in Fig.~\ref{fig:massessymmbreaking} and indicated in the inset. The dashed gray line corresponds to $v_R\simeq1\times10^{15}$, which is not consistent with the PTAs and hence ruled out. The colored shaded regions indicate the sensitivity curves of present (solid boundaries) LIGO O3 \cite{KAGRA:2021kbb}, NANOGrav \cite{NANOGrav:2023gor} and future (dashed boundaries) LIGO O5, SKA \cite{Smits:2008cf}, THEIA \cite{Garcia-Bellido:2021zgu}, LISA \cite{Baker:2019nia}, $\mu$-ARES \cite{Sesana:2019vho}, BBO \cite{Corbin:2005ny}, U-DECIGO \cite{Yagi:2011wg, Kawamura:2020pcg}, CE \cite{Reitze:2019iox} and ET \cite{Punturo:2010zz} experiments.}
    \label{fig:omegaftype1}
\end{figure}

In Fig.~\ref{fig:massessymmbreaking}, we have shown the neutrino mass evolution in the split seesaw mechanism versus the symmetry-breaking scale.  The colored shaded region indicates the NANOGrav \cite{NANOGrav:2023gor} excluded regime as illustrated in Fig.~\ref{fig:omegaftype1} via frequency bins. The inside colored curves indicate the different values of the exponential suppression factor. The colored stars correspond to the atmospheric scale mass, $M_\nu\simeq0.05$~eV for each exponential suppression factor for the atmospheric scale mass. The colliders constrain the lower value $v_R\gtrsim10^5$~GeV \cite{Queiroz:2024ipo} presented by the gray shaded regime.
It is worth noting that in the split seesaw mechanism, though the smallness of neutrino mass is gained by the exponential suppression of Yukawa couplings as given in \cref{eqn:seesawmassformula} but since the breaking scale $v_R$ is observationally constrained for a given value of $\beta$ it puts an upper bound on the suppression factor as shown in the Fig.~\ref{fig:massessymmbreaking}. Therefore, the formation of CSs in LRSM combined with neutrino data, constrains the full mass parameter space of the model.
\section{\textit{Type-I} Cosmic Strings}
\label{sec:type1cs}
The parameter $\beta = \lambda/(2g^2)$ critically determines the nature of the CSs. For $\beta<1$, the strings are of \textit{type-I} and exhibit attractive self-interactions, in contrast to the repulsive interactions of \textit{type-II} strings ($\beta>1$). As a consequence of this attraction, higher winding number bound states can be formed. In particular, these bound states prevent an ordered Abrikosov lattice from existing in laboratory experiments involving \textit{type I} superconductors \cite{Bettencourt_1995}. We refer the reader to Ref.~\cite{Ringeval:2005kr} for detailed analysis on the evolution of CS loops and to Refs.~\cite{Yung:1999du,Shifman:2002yi} for detailed analysis on \textit{type-I} and \textit{II} CSs.
A key feature of our framework is the natural emergence of \textit{type-I} CSs, characterized by $\beta < 1$. This is achieved through a small quartic coupling $\lambda$, which is well-motivated within the split seesaw mechanism. In conventional $4$D seesaw models, the small neutrino Yukawa coupling often requires fine-tuning. In our higher-dimensional setup, the small effective Yukawa is generated geometrically by wavefunction suppression, Eq.~\ref{5DYukawaSupression}. This liberates the quartic coupling $\lambda$ from the obligation to be large. In fact, a small value of $\lambda$ is technically natural. This naturalness is explicitly realized in supersymmetric (SUSY) extensions. In SUSY theories, the quartic couplings of scalar fields are not free parameters but are determined by the gauge couplings, leading to a natural expectation of $\lambda \sim O(0.1)$ \cite{Martin:1997ns}. While our present construction is non-SUSY, it is motivated by the fact that a small $\lambda$ is a robust feature of well-motivated UV completions. This distinguishes our scenario from a finely-tuned one and provides a strong theoretical motivation for exploring the $\beta < 1$ regime.

\begin{figure}[t]
    \centering
    \includegraphics[width=\columnwidth]{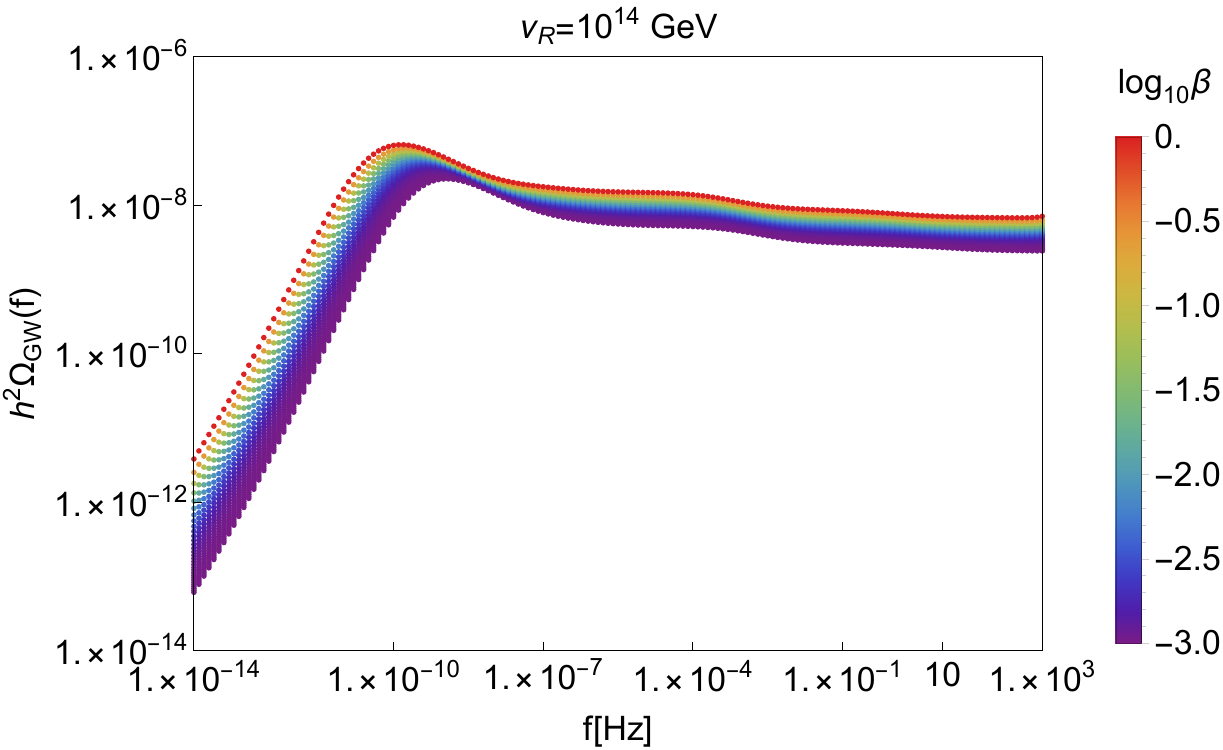}
    \caption{The evolution of GW spectra from CSs for different values of $\beta$ as shown by the vertical colored bar. We fix $v_R=10^{14}$~GeV, the amplitude of the GW spectrum decreases logarithmically with the decrease in $\beta$.}
    \label{fig:beta_var}
\end{figure}

Furthermore, the value of $\beta$ profoundly impacts the CS network evolution and its GW signature. While the expression for $\Omega_\text{GW}(f)$ in Eq.~\ref{GWdensity} has a generic form, it depends critically on the string tension $\mu_{cs} \propto v_R^2\, B(\beta)$. The function $B(\beta)$ varies from $1$ for $\beta \ll 1$ to $ln(\beta)$ for $\beta \gg 1$, leading to a significantly different normalization of the GW spectrum for the same symmetry-breaking scale $v_R$.
More importantly, the parameter $\beta$ determines the fraction of energy lost to gravitational radiation versus particle emission, which influences the loop number density (parameterized by $\zeta$ and $\alpha$ in Eq.~\ref{GWdensity}) \cite{Ringeval:2005kr, Blanco-Pillado:2011egf}. \textit{Type-I} strings have attractive self-interactions, leading to a different loop-chopping efficiency and network scaling solution compared to the more commonly studied \textit{type-II} case. Our analysis, which focuses on the $\beta \ll 1$ regime, therefore explores a distinct and observationally viable corner of the CS parameter space that is intimately linked to the neutrino mass generation mechanism. We have shown the impact of $\beta$ on the GW spectrum in Fig.~\ref{fig:beta_var} for a fixed value of $v_R=10^{14}$~GeV. The GW spectrum decreases logarithmically with the decrease in $\beta$.
Our analysis is now extended to the case $\beta\simeq0.01$ to put the constraints on the parameter space for a wide range of GW detectors.
\subsection{Phenomenological Implications}
\label{sec:phenomenCS}
Let us now explain the GW power spectrum for \textit{type-I} CSs. We follow the treatment given in \cite{Cui:2007js}. The total GW density at the present frequency $f$ is,
\begin{align}
    \label{GWdensity}
    \Omega_{\rm GW}(f)&\simeq \dfrac{2\,\Gamma\, G\mu_{cs}^2}{\rho_c\,f}\,\dfrac{\zeta}{\alpha\left(\alpha+\Gamma\,G\mu_{cs}\right)}\times \int_{t_f}^{t_0} d\tilde{t}\left(\dfrac{a(\tilde{t})}{a(t_0)}\right)^5\\\notag
    &\left(\dfrac{a(t_i)}{a(\tilde{t})}\right)^3\, t_i^{-4}.
\end{align}
Here $\Gamma\simeq 50$ quantifies the efficiency of GW emission, $G\mu_{cs}$ is the dimensionless string tension,  $\rho_c=3H_0^2/(8\pi G)$ is the critical density ($H_0$ is the Hubble parameter today and $G$ is Newton's constant). We considered the largest width of the loop parameterized by $\alpha=0.1$ and $\zeta=10$, which quantifies the net energy flux associated with the loops. The upper limit on the integration is up to the present time $t_0$, the lower limit $t_f$ is the time when the cosmic strings enter the scaling regime, $a(t)$ being the scale factor and $t_i$ is the loop formation time given as,
\begin{align}
t_i=\dfrac{\tilde{l}+\Gamma\,G\mu_{cs}\,\tilde{t}}{\alpha+\Gamma\,G\mu_{cs}}.
\end{align}
Here $\tilde{l}=2 a(\tilde{t})/(f a(t_0))$ is the size of the loop. We have presented the GW power spectra in Fig.~\ref{fig:omegaftype1}, which indicates the testability of the model and the symmetry-breaking scale in the present and future planned experiments. The different colors in the power spectrum indicate the benchmark point for the symmetry-breaking scale as indicated in Fig.~\ref{fig:massessymmbreaking}. To be consistent with the current PTA dataset \cite{NANOGrav:2023gor}, the symmetry-breaking scale $v_R\lesssim 10^{15}$~GeV as shown by the gray dashed line in Fig.~\ref{fig:omegaftype1}. Therefore, the emission of GWs from \textit{type-I} CSs provides testable signatures of neutrino masses through cosmology.
\begin{figure}[tbh]
    \centering
    \includegraphics[width=\columnwidth]{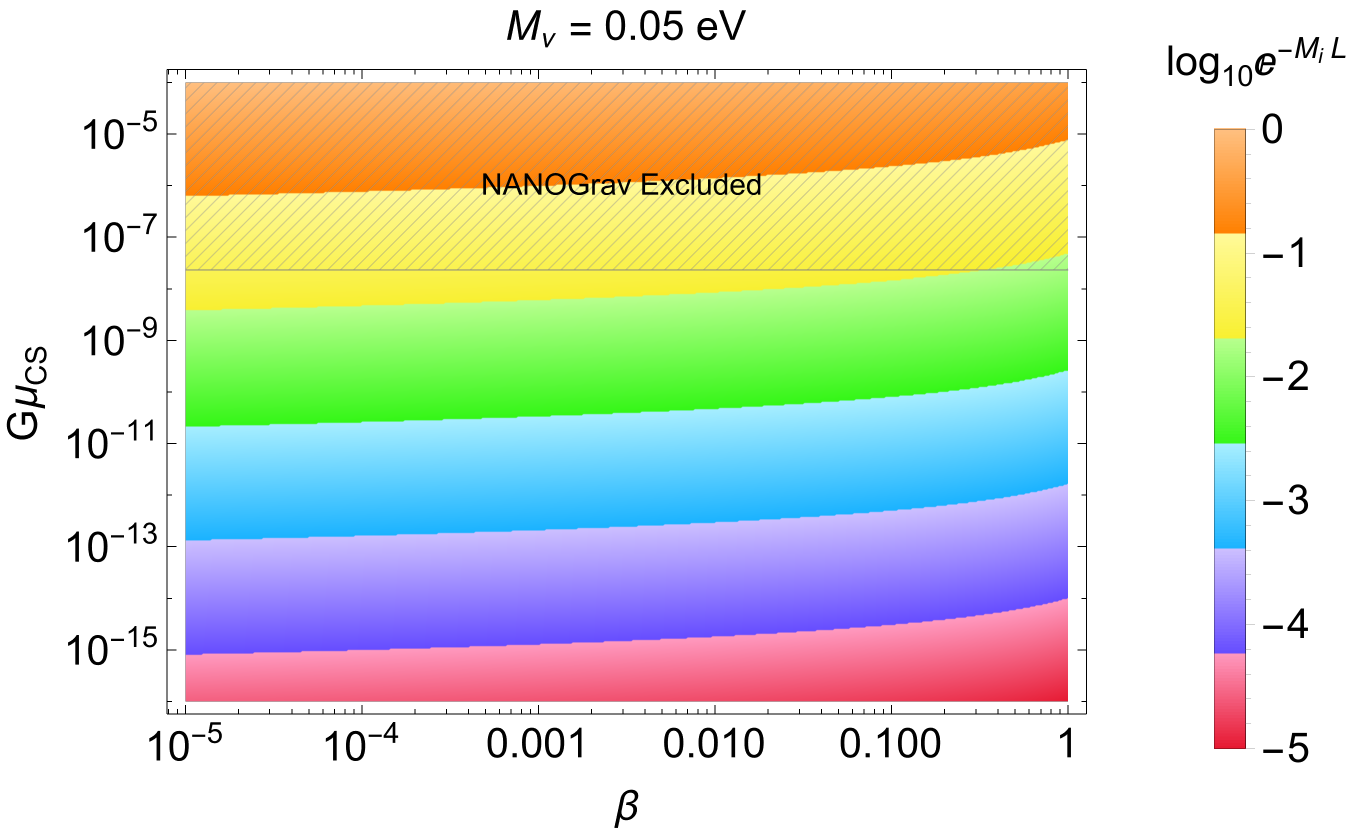}
    \caption{The evolution of the CS tension with $\beta$ and its impact on the exponential parameter appearing in Eq.~\ref{eqn:seesawmassformula} is shown with the vertical colored bar.
    The hatch region indicates the NANOGrav \cite{NANOGrav:2023gor} excluded regime. The mass of the neutrino is assumed to be the atmospheric scale mass, $M_\nu\simeq0.05$~eV.}
    \label{fig:summaryplot}
\end{figure}

It is worth noting that in the split seesaw mechanism, though the smallness of neutrino mass is gained by the exponential suppression of Yukawa couplings as given in Eq.~\ref{eqn:seesawmassformula}, since the breaking scale $v_R$ is observationally constrained for a given value of $\beta$ put an upper bound on the suppression factor as shown in the Fig.~\ref{fig:massessymmbreaking}. The summary of the constrained parameters are shown in Fig.~\ref{fig:summaryplot}. Where the evolution of the CS tension with beta has been presented and its impact on the exponential suppression factor appearing in Eq.~\ref{eqn:seesawmassformula} is shown with the vertical colored bar. The hatch-shaded region shows the NANOGrav excluded region. The LIGO O3 bound are less stringent than PTAs and the collider bound correspond to the very small value of $G\mu\sim 10^{-28}$.
\section{Conclusions}
\label{Con}
This work adopts the left-right symmetric model that provides a natural gauge-invariant embedding of right-handed neutrinos as part of $SU(2)_R$  doublets. By incorporating the split seesaw mechanism into this framework, one achieves exponentially suppressed effective Yukawa couplings due to wavefunction localization in extra dimensions, while the heavy Majorana mass $M_N$ arises from the spontaneous breaking of $SU(2)_R\times U(1)_{B-L}$. This unified picture not only explains the smallness of neutrino masses without fine-tuning but also predicts the formation of \textit{type-I} cosmic strings with $\beta \ll 1$ associated with the $B–L$ breaking scale. We have particularly shown that the suppression factor in the split seesaw mechanism is constrained by the choice of $\beta$ giving rise to the \textit{type-1} CSs. This framework addresses the neutrino mass generation from the split seesaw and its dependence on GWs emitted by \textit{type-1} CSs. In the split seesaw mechanism, one can explain the matter-antimatter asymmetry of the universe, as well as dark matter, which we leave for future work, but some discussion has already been made in Ref.~\cite{Kusenko:2010ik}.
\section*{Acknowledgments}
The author gratefully acknowledges Qaisar Shafi, Rishav Roshan, Rinku Maji, and Anish Ghoshal for their valuable discussions regarding CSs. Special thanks are also due to the anonymous referees for their insightful comments and suggestions, which significantly improved the manuscript.
\FloatBarrier
\bibliographystyle{apsrev4-1}
\bibliography{bibliography}
\end{document}